\documentclass[a4paper,12pt]{article}

\usepackage{amsmath,amsfonts,amssymb,epsfig}
\usepackage{enumerate}
\numberwithin{equation}{section}
\usepackage{comment}
\usepackage{cite}

\def\beq{\begin{equation}}
\def\eeq{\end{equation}}

\def\2b2[#1,#2][#3,#4]{\left( \begin{array}{cc} #1 & #2 \\#3 & #4 \end{array} \right)}
\def\3b3[#1,#2,#3][#4,#5,#6][#7,#8,#9]{\left( \begin{array}{ccc} #1 & #2 #3 \\#4 & #5 & #6\\#7&#8&#9\end{array} \right)}

\author{Karim~Benakli \footnote{kbenakli@lpthe.jussieu.fr} }
\date{}
\title{Disordered Extra Dimensions}
\begin{document}
\maketitle
\vspace{-1cm}
\begin{center}
\emph{Laboratoire de Physique Th\'eorique et Hautes Energies,  CNRS, UPMC Univ Paris 06
Boite 126, 4 Place Jussieu, 75252 Paris cedex 05, France}
\end{center}
\abstract{ A very large extra dimension may contain many  localized branes. We discuss the possibility to formulate such models as a spin system where each spin indicates  the supersymmetry direction preserved by the corresponding brane. In the  evolution of the universe, the extra dimensions might have ended into a  vacuum made of  patches with  different orientation of the spins, responsible for the observed breaking of supersymmetry. We discuss  the limit where the separation of these patches is by very thin defects  described as localization of  gravitino masses.}

\section{Introduction}

It might be useful, in order to account for the complexity of the world,  to embed  the observed world in a higher-dimensional space. This allows, by proper engineering, to  give a geometric origin for the low energy features of models. Extra dimensions have thus been introduced  about a century ago  by Nordstr\"{o}m, Kaluza and Klein \cite{Nordstrom:1988fi}.  Such a program has then been pursued  with a renewal of interest in the last decades (see for example \cite{Antoniadis:1990ew,ADD,DDG,Intermediate,RS,PQ}). Of particular interest for us, there are many attempts to engineer supersymmetry (SUSY) breaking mechanisms, classify them, and obtain experimental predictions. In these studies, the importance of extra dimensions  depends on the relative sizes of the supersymmetry breaking scale ($M_{SUSY}$) and the compactification one ($M_C$).  For $M_{SUSY}\ll M_C$, one can restrict the analysis to the four-dimensional effective theory (for a review, see for example \cite{Kitano:2010fa}). There, a knowledge of the data of the  extra dimensions  allows to "understand" the  fields content and their interactions.  In contrast, for $M_{SUSY}\gtrsim M_C$, the analysis should be performed in the  higher dimensional theory.  We are interested in the latter case.

A way of breaking supersymmetry with extra-dimensions is the Scherk-Schwarz  mechanism \cite{Scherk:1978ta},  where different
higher dimensional supersymmetries are conserved at different points. For instance, in five dimensional  models
compactified on an interval, $\mathcal{N}=2$ SUSY is preserved in the bulk, while at each of the two boundaries a different $\mathcal{N}=1$
supercharge survives, leading to supersymmetry  fully broken in the four-dimensional effective theory (see for example \cite{Antoniadis:1990ew,PQ,ADPQ,DPQ,Barbieri:2000vh,MP,Diego:2006py}). Here, we shall be interested in a straightforward  generalization of such a scenario, where instead of two branes at the boundaries, one deals with many branes at different points of the extra dimension(s).

Assumptions on the size, shape, and content of extra dimensions, turn in fact into assumptions on the very early history of the universe.
As the   extra dimensions  are expected to have a small volume, their evolution can be thought to be short, ending before nucleosynthesis to avoid unobserved variations of fundamental constants. Even if, for such a small size,  all parts are causally related, we want to argue that in the presence of branes the homogenization of this space might not have been efficient enough. In an alternative to the usual scenario of a symmetric internal space, we assume that the cooling  has been very fast, may be in a  non-adiabatic way, and as a consequence, the  internal space was not  driven all to a single ground state. Instead, we will assume that a situation  similar to the domains of ferromagnets  might arise.

In section 2, we describe the scenario of branes represented as a disordered system of spins living in the internal dimensional space. This is to be contrasted with the usual wisdom of considering symmetric internal spaces where, in a brane-world framework, the disorder consists in the presence of a single anti-brane as the source of SUSY breaking. In the disordered extra dimension set-up discussed here,  the "defects" separating different supersymmetric sub-spaces play a major role, as they are responsible for the breaking of (super)symmetry. Section 3, discusses the case of one extra-dimension, and illustrates how a "domain wall" can be approximated as a localization of a gravitino mass making further computations simpler. In fact, effects of localized gravitino masses are well studied in one extra dimension for the case of an interval with two boundaries\cite{Bagger:2001qi,Meissner:2002dg,Delgado:2002xf}, or  including many branes\cite{Benakli:2007zza}. In contrast, the case of localized mass in six dimensions has not yet been discussed to our  knowledge, and it will be treated in section 4. Most of the content of these two sections is original material that can be read independently of the rest. We end the paper by some conclusions.


\section{The world as a lattice and the branes as   spins}

 As for the observable ones,  properties of the extra dimensions, such as geometry and topology,  could be determined by the very early history of the universe.  Contrary to the large observed space-time,  very little is known about the evolution of the (small volume) internal space; some geometrical parameters have to be frozen  before nucleosynthesis as their values are associated with those of fundamental constants, whose variations are very constrained. The interaction between different parts of the extra dimensions can be achieved very quickly ($ t< 10^{-13}$s), as all parts can be causally connected. The  possibility that some degrees of freedom in the internal space are at a finite and sizable temperature is not excluded.
  
 The history of the internal space is assumed to proceed through three steps (i) space-time is nucleated (ii) diverse bubbles are created in the extra-dimensions where the branes are in excited states preserving no supersymmetry (iii)  while the "temperature" decreases, the branes move to minimize their energy  and different patches have branes that are ``pointing'' towards a different supersymmetry. The internal space finishes  frozen in a (meta-stable) state  made of  "domains" described by different ground states.

The branes can be located at arbitrary points in the extra-dimensions. Patches in well defined supersymmetric ground states  have  branes positioned as long molecules of a liquid crystal in a nematic phase. One can also consider the branes on a regular lattice, the inter-branes spacing is then  fixed by some kind of Van der Waals forces  \cite{ArkaniHamed:1998kx}, or by the equilibrium between forces due to a combination of electric and dyonic charged branes  \cite{Corley:2001rt,Denef:2000nb}. Such mechanisms appeal to physics at scales of order or smaller than the inter-branes separation, which needs the knowledge of the field content of the fundamental theory. The  existence of such a possibility to locate the branes at fixed position would  imply the non-supersymmetric vacuum to be (meta)stable, and also stabilizes the size of the extra-dimension \cite{ArkaniHamed:1998kx}. As we stress again in the conclusions,  achieving the stability of non-supersymmetric configurations is not obvious, and remains an open issue in string theory.

For each brane $i$,  we associate a vector, we will denote as the spin $\vec{S_i}$, corresponding to the central charge in the $\mathcal{N}=2$ super-algebra which describes the direction of the supersymmetry preserved by the brane. The spins $\vec{S_i}$, with  unit norm  $|\vec{S_i}|^2=1$, live in a two-dimensional space spanned by the unitary orthogonal basis vectors $(\vec{e}_X, \vec{e}_Y)$:
\begin{equation}
\vec{S}_i = \cos{\theta_{i}}\  \vec{e}_X + \sin{\theta_{i}} \ \vec{e}_Y
\label{Hamilton1}
\end{equation}
and we will define $\vec{e}_Z=\vec{e}_X \wedge \vec{e}_Y$.

 While, in the extra-dimensions discussed here, the branes appear as localized spins at some points, in  other smaller dimensions they are wrapping cycles  intersecting at angles that define the associated spin. In treating the branes as a spin system, one can have: i)  long range interactions (This is the case for toroidal constructions in string models) ii) the  values for the spins  to take discrete values.  Because of these, there would be no  fluctuation destruction of long-range order following the Mermin-Wagner-Berezinskii theorem, and supersymmetry ordering is expected in all dimensions.   Moreover, the spins system is supposed to be nowadays at zero temperature. Unless stated otherwise,\textit{ we will always implicitly assume the spins to take discrete values} but we shall consider finite range spin-exchange interactions.

Working in an effective description, we can take for the  Hamiltonian of this system the very simple form:
\begin{eqnarray}
\mathcal{H}&=& \sum_{i,j} J_{ij} |\vec{S}_i \wedge \vec{S}_j|^p+ \sum_{i}a_i |\vec{S}_i \wedge \partial_t \vec{S}_i|^q - \sum_{i} \vec{H}_i \cdot  \vec{S}_i \\
&=& \sum_{i,j} J_{ij} |\sin{(\theta_{j}-\theta_{i})}|^p+ \sum_{i} a_i |\partial_t {\theta}_i|^q -\sum_{i} b_i \cos({\theta}_i-{\theta}^{(H)}_i)
\label{Hamilton2}
\end{eqnarray}
where we have used $\vec{H}_i = b_i (\cos{\theta^{(H)}_{i}}\  \vec{e}_X + \sin{\theta^{(H)}_{i}} \ \vec{e}_Y)$. Here, $\{p,q\}$ are exponents that depend on the microscopic theory, we take to be  $p=q=2$ .

Let us discuss each term:
\begin{itemize}
\item The first term tends to align all the spins, giving a supersymmetric ground state with all the branes preserving the same $\mathcal{N}=1$ supersymmetry. The interaction strength $J_{ij}$ are  positive with  strength  and action range depending on the microscopic model.

\item The second term is included to take into account that the rotation of the spin requires a rotation of the brane which costs a non-zero amount of energy. This is because the supersymmetry preserved by the brane is associated to its orientation in other smaller extra dimensions. The rigidity $a_i$ could be related to the brane tension, and it does not need to be the same at each point.

\item The third term describes the effect of the background fields. It will determine the non-trivial vacuum structure as it leads to  constraints that pull the branes towards a specific direction. This can contain, for example, the effect of the presence of specific boundary conditions in the internal space  (as orientifold objects). 

\end{itemize}

It is the last term of Eq. (\ref{Hamilton1}) that parametrizes the source of supersymmetry breaking. Our aim then would be to find simple forms of $ \vec{H}_i$ that, on one side, lead to a minimum of the Hamiltonian with broken supersymmetry, and on the other side, can  {\it a posteriori} be understood at the level of a microscopic fundamental theory. 

The system of $N$ spins with constrained boundaries can be thought to evolve during the early time  from the initial conditions as

\begin{equation}
\frac{\partial \vec{H}_i}{\partial t}=\vec{F}_i(\vec{H}_j, \vec{S}_j ) \qquad \mbox{with    }  \qquad  \vec{H}_0(t)=\vec{H}_0, \qquad \vec{H}_N(t)=\vec{H}_N
\label{b2g}
\end{equation}
where the exact form of $\vec{F}_i$ requires the knowledge of the microscopic theory. A simple phenomenological approximate of such  equation could take the form
\begin{equation}
\vec{F}_i(\vec{H}_j, \vec{S}_j )=\sum_{j}\alpha_{ij} \vec{H}_j\wedge \vec{H}_i + \sum_{j}\beta_{ij} \vec{S}_j \wedge \vec{H}_i
\label{b2f}
\end{equation}
that tend to bring the vacuum to aligned spins but for the imposed non-supersymmetric boundary conditions. With  appropriate $\alpha_{ij},\beta_{ij}$, one  can achieve a background made of domains preserving different supersymmetries.

As stated above, the problem of supersymmetry breaking is now formulated as the problem of obtaining appropriate forms of the constraints  $ \vec{H}_i$ and study their properties as the phase space and the correlation function in the corresponding  spin model. 

Examples of constraints that allow obtaining ground states with domains where the spins point towards different directions can be constructed either by having a strong localized force, or a weak long range force, in both cases opposing the effect of the first term of (\ref{Hamilton1}) . We will illustrate this through examples for one extra dimension. 

 The first option  can be realized as:
 \begin{equation}
 \vec{H}_i=\sum_I b_I    \exp[{-\frac{(y_i-Y_I)^2}{\Delta}}](f_I(y)\vec{e}_X + g_I(y)\vec{e}_Y); \qquad b_I \gg J_{ij}
\label{b22}
\end{equation}
where $f_I,g_I$ are slowly varying functions that force the spins to change directions around the point $Y_I$. For example, if the points $Y_I =0, \pi R$ are the boundaries of a compact dimension, we can take:
\begin{equation}
  \vec{H}_i= b_0    \exp[{-\frac{(y_i)^2}{\Delta}}]\vec{e}_X + b_\pi   \exp[{-\frac{(y_i-\pi R)^2}{\Delta}}] \vec{e}_Y
\label{b2b}
\end{equation}
Another example is used in the next section.

The second option can also be obtained from the interplay of weak and long range forces. For instance, the aligning force is taken to be effective only between nearest neighbors, while parametrizes an anti-alignment force at long range:
\begin{equation}
 \vec{H}_i=\sum_{j\neq i}    \frac{b}{(y_j-y_i)^\alpha} \vec{S}_j
\label{b2c}
\end{equation}
where $\alpha$ is a positive integer. We take $J_{ij}$ to a nearest neighbor interaction, and  $b$ positive and small  $b\ll J$, we see that the energy of the system is lowered for spins aligned in domains, but a large amount of such spins generate an interaction that tends to flip the spins. The net result is the formation of domains with different alignments separated by transition regions.

 In this picture of disordered extra dimension, the supersymmetry breaking is localized  in the "defects" separating different supersymmetric sub-spaces. It is the subject of the next sections. We will assume that the fundamental length scale $\kappa$, the inter-brane separation $d_i$ and the compactification scale $R$ are well separated $\kappa \ll d_i\ll R$. The first inequality allows to neglect quantum geometry effects, the second to get a large number of branes. When needed for the purpose of explicit computations, we will also use a flat metric.


\section{A defect in one extra dimension}

We consider the space-time extended with a fifth dimension with coordinate $y \in [0, \pi R]$. Along this direction $N + 1$ branes are located at the points $y = y_{i}$, $i = 0 \cdots N$ with $y_{0} = 0$, $y_{N} = \pi R$, and $y_{n} < y_{n+1} $. 

 We assume that the evolution of the universe has ended in a non-supersymmetric vacuum. The branes orientation varies when going from one to the opposite boundary, such that the distribution of the associated spins has the form of a localized kink.   Obviously, non-trivial boundary conditions are needed  such that at different regions the (spin) branes have to point to different directions.  An abrupt separation of domains would be costly in spin-exchange energy. In fact, with the combination of boundary conditions, spin-exchange, and long range interactions, the minimization of the total energy will induce a kink with a certain thickness $\Delta'$ localized around a point $Y$, as for ferromagnets. As translation invariance is broken by both the presence of boundaries and  the conditions imposed  to break of supersymmetry, the minimization of the total energy will fix the value of $Y$ (i.e. gives masses to the collective modes of the soliton). Because, we are not interested here in this issue of fixing the moduli, we choose a final configuration, and illustrate how it can be parametrized as the result of an applied ``magnetic field'' $H_i$ on the spin system. Our aim will be  to show how our order parameter, the gravitino mass, is related to the final spin configuration.

We will first show how a kink localized at a single position $Y$ can be described as if the branes are embedded in a background field $\vec{H}_i$.  We consider the constraint: 
\begin{equation}
 \vec{H}_i=b    \exp[{-\frac{(y_i-Y)^2}{\Delta'}}]( \cos{(\frac{y_i}{\Delta})} \vec{e}_X + \sin{(\frac{y_i}{\Delta})} \vec{e}_Y)
\label{b2d}
\end{equation}
which means that at the boundary $y=0$, the spins at forced to point in the direction $\vec{e}_X $ while at $y=\pi R$, the spins at forced to point along $\vec{e}_Y $, as well as 
\begin{equation}
 J_{i j} = J[ \delta_{j,i+1}+ \delta_{j,i-1}]; \qquad a_i=a
\label{J2}
\end{equation}
The Hamiltonian is given by:
\begin{equation}
\mathcal{H}= \sum_{i} J \sin^2{(\theta_{i}-\theta_{i+1})}+ \sum_{i} a |\partial_t {\theta}_i|^2 -\sum_{i} b \exp[{-\frac{(y_i-Y)^2}{\Delta'}}] \cos{( \theta_i-\frac{y_i}{\Delta})} 
\label{H2}
\end{equation}

The minimum of the potential is obtained for:
\begin{equation}
 \sin{2(\theta_{i}-\theta_{i+1})}= -\frac{b}{J} \exp[{-\frac{(y_i-Y)^2}{\Delta'}}] \sin{( \theta_i-\frac{y_i}{\Delta})} 
\label{min1}
\end{equation}
Taking $b\gg J$, we obtain the following limits:

In the vicinity of   Y,  $y_i \sim Y$, we can approximate  
\begin{equation}
\sin{( \theta_i-\frac{y_i}{\Delta})} = -\frac{J}{b} \sin{2(\theta_{i}-\theta_{i+1})} \rightarrow 0  \qquad  \mbox{i.e.    }   \qquad  \theta_i \rightarrow \frac{y_i}{\Delta}
\label{min2}
\end{equation}
while elsewhere:
\begin{equation}
 \sin{2(\theta_{i}-\theta_{i+1})} \rightarrow 0  \qquad  \mbox{i.e.    }    \qquad  \theta_{i}  \rightarrow \theta_{i+1}
\label{min3}
\end{equation}
This means we have  two patches of aligned spins, separated by a region of size of order $2 \Delta'$. In his transition region, the spins rotate by an angle of order $2  \Delta'/ \Delta$, which if not a multiple of $\pi$, implies that the two patches preserve two different supersymmetries. It is only this quantity that will be relevant for our purpose. Of course, one can compute the energy carried by the interface between the two domains, and study the process of bubbles nucleation, the size of the interface, or the processes of homogenization, all well known issues.

We are interested to describe here the effect of the simplest such defect  in a five-dimensional supergravity. The total action is given by the sum of a bulk and brane components:
\begin{equation}
    S =  \int^{2 \pi R}_{0} dy \int d^{4}x \left[ \frac{1}{2} {\cal L}_{BULK} + 
\sum_{i = 0}^{N} {\cal L}_{i} \delta({y - y_{i}})  \right] .
\label{ActionN}
\end{equation}

The brane $n$ will be characterized by the supersymmetry it preserves, which is correlated with the choice of the couplings to the bulk operators, in particular the gravitino. The non-vanishing set of such operators $\Phi_{even}$ are  determined as those being  even  under a $\mathbb{Z}_2$
action at the point $y = y_{i}$:
\begin{equation}
\Phi_{even}(y_{i} + y) = {\cal P}_{i} \Phi_{even}(y_{i} - y)=
\Phi_{even}(y_{i} - y).
\label{ParityN}
\end{equation}
The operators might be themselves made of products of even numbers of odd fields. The supersymmetry preserved by the brane $i$, associated with the ``spin'' $\vec{S_i}$ can be read from the gravitino components $\psi_{\mu +}^{~i}$ which couples to it, while it breaks the orthogonal supersymmetry direction associated to $\psi_{\mu -}^{~i}$. We can choose a basis $(\psi_{\mu 1} , \psi_{\mu 2})$ for the $\mathcal{N}=2$ gravitino in term of two-components spinors, and 
define:
\begin{eqnarray}
\psi_{\mu +}^{~i} & = &
\cos{ 2\theta_{i} } \  \psi_{\mu 1} + \sin{ 2\theta_{i} } \  \psi_{\mu 2} 
\nonumber \\
\psi_{\mu -}^{~i} & = &
 \! \! \!  \! \! - \sin{2 \theta_{i} }\   \psi_{\mu 1} + \cos{2 \theta_{i} } \  \psi_{\mu 2} 
\nonumber \\
\psi_{5 +}^{~i} & = &
\sin{2 \theta_{i} }\   \psi_{5 1} + \cos{ 2\theta_{i} } \  \psi_{5 2} 
\nonumber \\
\psi_{5 -}^{~i} & = &
\cos{2 \theta_{i} }\   \psi_{5 1} - \sin{2 \theta_{i} }\   \psi_{5 2} 
\label{ParityEigenvectorsN}
\end{eqnarray}
Without lost of generality, we will take $\theta_{0} = 0$. The spin $\vec{S_i}$ of the chain makes an angle $\theta_i$ with $\vec{S_0}$.

We shall now consider the case of a single domain wall separating two phases. the generalization to more than one domain is straightforward. At leading order, we shall consider the branes world-volumes to be supersymmetric. The whole supersymmetry breaking is then concentrated in the transition interval $\left[ y_n, y_{n+1}\right]$ of length $d_n$. It is encoded in the wave function of the gravitino which interpolates between the two values $\psi_{\mu +}^{~n}$ and $\psi_{\mu +}^{~n+1}$.   This variation is associated with a gravitino mass $M_{n} (\theta_n, d_n)$. We take a configuration where:
 \begin{itemize}   
 \item all  spins  $\vec{S_i}$, $0\leqslant  i \leqslant n$ are aligned with  $\vec{S_0}$, thus $\theta_{0} = ...= \theta_n=0$. We will denote this as  the phase $A$ .
 \item all spins  $\vec{S_i}$, $n+1\leqslant  i \leqslant N$ are aligned with  $\vec{S_N}$, thus $\theta_{n+1} = ...= \theta_N=2 \theta$. We will denote this as  the phase $B$. 
\end{itemize}

For the purpose of the illustration,  we can take the extra dimension to be flat, thus $M_{n} (\theta, d_n) = \theta/ d_n$. The gravitino wave function associated to the supersymmetry preserved on the left side of the defect is given by:
\begin{eqnarray}
\psi_{\mu 1}(y)   = \left\{
    \begin{array}{lll}
        1 & \mbox{for } y \in [0, y_{n}] \\
       \cos{ ( \frac{y-y_n}{d_n}2 \theta)} & \mbox{for } y \in [y_{n}, y_{n+1}] \\
       \cos{2 \theta} & \mbox{for}  y \in [y_{n+1}, \pi R]
    \end{array}
\right.
\end{eqnarray}
while the orthogonal one is given by
\begin{eqnarray}
\psi_{\mu 2}(y)   = \left\{
    \begin{array}{lll}
        0 & \mbox{for } y \in [0, y_{n}] \\
      \sin{ ( \frac{y-y_n}{d_n} 2\theta)} & \mbox{for } y \in [y_{n}, y_{n+1}] \\
      \sin{2 \theta} & \mbox{for } y \in [y_{n+1}, \pi R]
    \end{array}
\right.
\end{eqnarray}
such that the right side of the defect preserves the combination: $ \cos{ 2 \theta } \psi_{\mu 1} + \sin{2 \theta } \psi_{\mu 2} $.

For a bulk observer outside the domain $\left[ y_n, y_{n+1}\right]$, in the limit $d_n\ll \pi R$ the breaking of supersymmetry can be accounted a variation  of the bulk field (here the gravitino's) wave function between the points  $y_n$ and $y_{n+1}$. We would like to describe the wave function outside the defect in the limit where the latter can be considered as point-like, i.e $d_n \rightarrow 0$.

In this limit,  we  describe the five-dimensional gravitino by two wave functions: a continuous one $\Psi_{C}$ that couples to the defect with a mass $M_n$,   and a discontinuous one, $\Psi_{D}$, that does not couple. We can find the respective values building these functions in the interval $\left[ y_n, y_{n+1}\right]$:

\begin{eqnarray}
\Psi^{C n}_{\mu}(y) & = &  c_{\theta}  \psi^{n}_{\mu 1}(y)+ s_{\theta} \psi^{n}_{\mu 2}(y)\\
\Psi^{D n}_{\mu}(y) & = & s_{\theta} \psi^{n}_{\mu 1}(y)-  c_{\theta} \psi^{n}_{\mu 2}(y)
\end{eqnarray}
where:
\begin{eqnarray}
 c_{\theta}= \cos{  \theta}, && s_{\theta}= \sin{ \theta} 
\end{eqnarray}
In the limit $d_n \rightarrow 0$,  $y_n = y_{n+1}= Y_n$,  the gravitino component that couples to the defect domain wall at $y_n = y_{n+1}= Y_n$ is given by the even wave function value:
\begin{eqnarray}
\Psi^{C n}_{\mu}(Y^<_n)= \Psi^{C n}_{\mu}(y_n)  = c_{\theta}= \Psi^{C n}_{\mu}(y_{n+1}) = \Psi^{C n}_{\mu}(Y^>_n)
\end{eqnarray}
while the orthogonal component 
\begin{eqnarray}
\Psi^{D n}_{\mu}(y_n) & = &s_{\theta}= - \Psi^{D n}_{\mu}(y_{n+1})
\end{eqnarray}
is odd and corresponds to the a gravitino component that does not couple to the defect. 

We can now build, in this limit,  an  effective wave function:
\begin{eqnarray}
\Psi^{C }_{\mu}(y)  = \left\{
    \begin{array}{llll}
     c_{\theta}    & \mbox{for } y \in [0, Y_n] \\
         s_{\theta}  & \mbox{for }  y \in [Y_{n}, \pi R]
    \end{array}
\right.
\end{eqnarray}
orthogonal to
\begin{eqnarray}
\Psi^{D }_{\mu}(y)   = \left\{
    \begin{array}{lll}
       c_{\theta}& \mbox{for } y \in [0, Y_{n}] \\
    -  s_{\theta} & \mbox{for } y \in [Y_{n}, \pi R]
    \end{array}
\right.
\end{eqnarray}

The breaking of supersymmetry by  defect can now be described as due to a localized mass term $M_n$. Such a localized mass gives rise to the equations of motion for the gravitinos $\Psi^{I  n}_{\mu}$(we assume $e^{\hat{5}}_{5} =
1$):
\begin{eqnarray}
     \partial_{5}\Psi^{D }_{\mu} 
+ m_{3/2} \Psi^{C }_{\mu }
&=& 2 M_n \Psi^{C }_{\mu} \delta(Y_n)
\nonumber    \\
     \partial_{5} \Psi^{C}_{\mu} 
- m_{3/2}\Psi^{D}_{\mu}
&=& 0
\label{GravitinosEOM2}
\end{eqnarray}
where we have used the four-dimensional equation of
motion for  gravitinos  of mass $m_{3/2}$:
\begin{equation}
\epsilon^{\mu \nu \rho \lambda}
\sigma_{\nu}\partial_{\rho}\overline{\Psi}^{I}_{\lambda } = 
- 2 m_{3/2} \sigma^{\mu \nu} \Psi^{I }_{\nu } \qquad I=C, D
\label{MassiveGravitinoEOM}
\end{equation}

It can be clearly seen from equations (\ref{GravitinosEOM2})  that while $\Psi^{C }_{\mu}$ is a 
continuous field,  $\Psi^{D i}_{\mu }$   has a 
jump at the point $y = Y_n$, its first derivative being proportional to a
Dirac $\delta$ distribution. We can then identify the gravitino mass as:
\begin{equation}
M_n =  \kappa^{-1} \tan{  \theta_{n+1}} = \kappa^{-1} \frac {(\vec{S}_{n+1}\wedge \vec{S_{n}}) \cdot \vec{e}_Z }{(\vec{S}_{n+1}\cdot \vec{S_{n}})}
\label{MassiveGravitino}
\end{equation}
Given the knowledge of the localized gravitino mass in 5d, we can use this result to derive the  four-dimensional gravitino mass.
\begin{equation}
m_{3/2} =  \frac{1}{\pi R} \arctan{( \frac {(\vec{S}_{n+1}\wedge \vec{S_{n}}) \cdot \vec{e}_Z }{(\vec{S}_{n+1}\cdot \vec{S_{n}})}})
\label{MassiveGravitino2}
\end{equation}
As an illustration of this simple case,  let us consider  $\kappa \lesssim d_n \sim$ TeV$^{-1}$. The observable sector lives on a 4-brane extended between the two points $y_n$  and $y_{n+1}$, that is part of a large extra dimension of size $\pi R$ responsible for the hierarchy between the string and the Planck length. We can use the previous example to see that the resulting gravitino mass is $  \theta_{n+1}/\pi R$. In the case of a system of brane-anti-brane, $ \theta_{n+1}=  \pi/2 $ leads to $m_{3/2} = 1/2R$. As explained in \cite{Benakli:2007zza} for the case of an explicit localized F-term, the breaking can not be compensated by opposite twists in other parts of the extra dimension.


\section{A localized defect in two dimensions }
\label{secSixDSugra}

In this section, we will illustrate  the case  of supergravity with two extra dimensions, i.e. in six dimensions.  The defects can be either  a one-dimensional curve or a point. The latter can appear in the spin system as the zero size limit  of a vortex. The gravitino wave functions can be taken in the absence of supersymmetry breaking as holomorphic (or anti-holomorphic) function of the complex coordinate $z= x^5+ix^6$ describing the two extra dimensions. When a gravitino mass $m_0$ localized at the point $z_0$ is included for the component ${\psi}_{\mu 1}$, it appears as a flux in the circulation of the gravitno wave function ${\psi}_{\mu 2}$, of the form 
\begin {equation}
\oint_{\partial S} {\psi}_{\mu 2} dz = -i m_1 \int_S {\psi}_{\mu 1}  dx^5 dx^6 = - 2 im_0 {\psi}_{\mu 1}(z_0)
\label {GravitinoKineticTerm1}
\end {equation}
where $S$ is a surface containing the point $z_0$ and having as boundary $\partial S$, while $m_1$ is the bulk mass appearing in the equation of motion of ${\psi}_{\mu 1}$. In this section, we will derive  the resulting lightest four-dimensional  gravitino mass.

The two extra dimensions  are taken compactified   on the orbifold $ T^2 / \mathbb { Z }_2 $ parametrized by the  coordinates $(x^5 , x^6)$. The torus $T^2$ coordinates obey $(x^5 , x^6) \equiv (x^5 + 2 \pi m R_5, x^6 + 2 \pi n R_6)$, $(m,n)\in \mathbb{Z}$, and the orbifold is obtained through the identification  ${(x^5 , x^6)} \equiv {-(x^5 , x^6)}$. There are four fixed points of this action at $(0,0)$, $(\pi R_5,0)$, $(0,\pi R_6)$ and $(\pi R_5, \pi R_6)$.  We will consider the simplest case with a single defect located at the origin $(x^5 , x^6) = (0,0)$.

The bulk Lagrangian volume must describe the six-dimensional supergravity. The supermultiplets of supergravity in six dimensions in its minimal form are the sechsbein $ e^a_m$, the gravitino $ \Psi_m$,   a real scalar field $ \Phi $, the dilaton; a fermion $ X $  the dilatino; and the Kalb-Ramond two-form  denoted by $ B_{MN} $ which gives rise to the three-form $ H = 3 \partial_{[B_M {NP}]} $. The action of supergravity in the volume is $ N = 2 $ supersymmetric as it preserves eight supercharges. Our study focuses on the gravitino. Its standard kinetic term  reads:

\begin {equation}
{\cal L}_{kin} = - i E_6 M^2_6 \overline {\Psi}_M \Gamma^{MNP} D_{N} \Psi_{P}
\label {GravitinoKineticTerm}
\end {equation}
where $M_6 = \kappa^{-1} $ is the fundamental Planck mass in six dimensions,  $E_6 $  the sechsbein determinant. It is useful to express this in  two-components spinor notation:
\begin{eqnarray}
{\cal L}_{kin} &=& \kappa^{-2} e_6  \bigg[ 
\frac{1}{2} \epsilon^{\mu \nu \rho
\lambda} \left( 
\overline{\psi}_{\mu 1}\overline{\sigma}_{\nu}D_{\rho}\psi_{\lambda 1}
+ \overline{\psi}_{\mu 2}\overline{\sigma}_{\nu}D_{\rho}\psi_{\lambda 2}
\right)
\nonumber    \\
&&
+ \psi_{\mu 1} \sigma^{\mu \nu} \left( D_{\hat{5}} + i D_{\hat{6}} \right)  \psi_{\nu 2} 
- \psi_{\mu2} \sigma^{\mu \nu} \left( D_{\hat{5}} + i D_{\hat{6}} \right) \psi_{\nu1} 
\nonumber    \\
&&
- \left( \psi_{\hat{5}1} + i \psi_{\hat{6}1} \right) \sigma^{\mu \nu}D_{\mu}\psi_{\nu2} 
+ \left( \psi_{\hat{5}2} + i \psi_{\hat{6}2} \right) \sigma^{\mu \nu}D_{\mu}\psi_{\nu 1} 
\nonumber    \\
&&
- \psi_{\mu 1} \sigma^{\mu \nu}D_{\nu} \left( \psi_{\hat{5}2} + i \psi_{\hat{6}2} \right)
+ \psi_{\mu 2} \sigma^{\mu \nu}D_{\nu} \left( \psi_{\hat{5}1} + i \psi_{\hat{6}1} \right) 
\bigg]+ h.c. 
\label{KineticTerm2}
\end{eqnarray}

To define this theory in the orbifold $ T^2 / \mathbb {Z}_2 $ we must impose parity of various fields under the action of the symmetry $ \mathbb{Z}_2 $ in a manner consistent with the action of supergravity and supersymmetry transformations. Expressing  $ \Psi_m $ and $ X $  in two-components spinor notation,  (i) the fields $ e^a_{\mu} $, $ e^{i_{j}} $, $ B_{\mu \nu} $, $ B_{ij} $, $ \Phi $, $ \psi_{\mu 1} $, $ \psi_{i 2} $ and $ \chi_1$ are taken even under  $ \mathbb{Z}_2 $ (ii) the fields $ e^i_{\mu} $, $ e^{a}_{i} $, $ B_{\mu i} $ $ \psi_{\mu 2} $, $ \psi_{i 1} $ and $ \chi_2$ are odd under the $ \mathbb {Z}_2 $ action. Here the indices $ i, j$  denote coordinates of the extra dimensions: $ i, j \in \{5,6\} $.

The defect is located at the fixed point of the orbifold $ (x^5, x^6) = {(0,0)} $,  therefore only the operators even under the  $ \mathbb {Z }_2 $ action couple to it.  We are interested in the case where  a constant localized  four-dimensional gravitino mass:
\begin {equation}
{\cal L}_{mass} = - e_4 \delta {(x^5)} \delta {(x^6)}
\left (M_0 \psi_{\mu 1} \sigma^{\mu \nu} \psi_{\nu 1}+ hc \right)
\label {BraneMassTerm}
\end {equation}
is present, as well as new bi-linear terms which mix the four-dimensional gravitino $ \psi_{\mu 1} $ with the internal dimensional components $ \psi_{5 2} $ and $ \psi_{6 2} $. The constant $ M_0 $ is proportional to the value  $ W_0 $ of the localized superpotential: $ M_0 = \sqrt {g_{\hat {5} \hat {5}}} \, W_0 $. A necessary step is gauge fixing. A possible choice  is the unitary gauge where the terms bi-linear mixing the four-dimensional gravitino $ \psi_{\mu } $ fields with $ \psi_{5} $ and $ \psi_{6} $ are absent, so that the part of the Lagrangian which describes the bi-linear terms for the gravitino is given by:
\begin{eqnarray}
{\cal L}_{k + m} &=& \kappa^{-2} \left[ 
\frac{1}{2} \epsilon^{\mu \nu \rho
\lambda} \left( 
\overline{\psi}_{\mu 1}\overline{\sigma}_{\nu} \partial_{\rho}\psi_{\lambda 1}
+ \overline{\psi}_{\mu 2}\overline{\sigma}_{\nu} \partial_{\rho}\psi_{\lambda 2}
\right)
+ 2 \psi_{\mu 1} \sigma^{\mu \nu} \left( \partial_{5} + i \partial_{6} \right)  \psi_{\nu 2} 
\right]
\nonumber    \\
&&
- \delta(x^5) \delta(x^6) M_0 \psi_{\mu 1} \sigma^{\mu \nu} \psi_{\nu 1}  
+ h.c.
\label{KineticAndMassTerms}
\end{eqnarray}

To study the properties of the gravitino there are two approaches: one can study its equations of motion and boundary conditions as done in the one-dimensional case, or we can study the theory reduced to four dimensions. In this section we follow the second method.

First, we Fourier expand the  gravitinos $ \psi_{\mu 1} (x^{\mu}, x^5, x^6) $ and $ \psi_{\mu 2} (x^{\mu}, x^5, x^6)$, taking into account their parities under the $ \mathbb {Z}_2 $ action:

\begin{eqnarray}
\psi_{\mu 1}(x^{\mu},x^5, x^6) &=& \frac{\kappa}{\sqrt{\pi^2 R_5 R_6 }} \left[ 
\frac{1}{\sqrt{2}}\psi_{\mu 1}^{0}(x^{\mu}) 
+ \sum_{p,q \in Y} \psi_{\mu 1}^{p,q}(x^{\mu}) 
\cos \left( \frac{p x^5}{R_5} + \frac{q x^6}{R_6} \right) \right] 
\nonumber    \\
\psi_{\mu 2}(x^{\mu},x^5, x^6) &=& \frac{\kappa}{\sqrt{\pi^2 R_5 R_6 }} 
\sum_{p,q \in Y} \psi_{\mu 2}^{p,q}(x^{\mu}) 
\sin \left( \frac{p x^5}{R_5} + \frac{q x^6}{R_6} \right) 
\label{FourrierExpansion}
\end{eqnarray}
with the sum over $Y$ is defined as: $\sum_{p,q \in Y} = \sum^{p = + \infty}_{p = 1}\sum^{q = + \infty}_{q = - \infty} +  \left[ \sum^{q = + \infty}_{q = 1} \right]_{{p = 0}} $ . When plugged in (\ref {KineticAndMassTerms}), it gives:

\begin{eqnarray}
{\cal L}_{k + m} &=&  \frac{1}{2} \epsilon^{\mu \nu \rho \lambda} \left[ 
\overline{\psi}^0_{\mu 1}\overline{\sigma}_{\nu}\partial_{\rho}\psi^0_{\lambda 1}
+ \sum_{p,q \in Y} \overline{\psi}^{p,q}_{\mu 1}\overline{\sigma}_{\nu}\partial_{\rho}\psi^{p,q}_{\lambda 1}
+ \sum_{p,q \in Y} \overline{\psi}^{p,q}_{\mu 2}\overline{\sigma}_{\nu}\partial_{\rho}\psi^{p,q}_{\lambda 2}
\right]
\nonumber    \\
&&
- \frac{M_0 \kappa^2}{\pi^2 R_5 R_6} \left[ 
\frac{1}{\sqrt{2}} \psi_{\mu 1}^0 + \sum_{k,l \in Y} \psi_{\mu 1}^{k,l} \right]
\sigma^{\mu \nu} \left[
\frac{1}{\sqrt{2}} \psi_{\nu 1}^0 + \sum_{p,q \in Y} \psi_{\nu 1}^{p,q} \right]
\nonumber    \\
&&
+ 2 \sum_{p,q \in Y} \psi^{p,q}_{\mu 1} \sigma^{\mu \nu} 
\left( \frac{p}{R_5} + i \frac{q}{R_6} \right) \psi^{p,q}_{\nu 2}
+ h.c.
\label{4dLagrangian}
\end{eqnarray}

Note that the phases in the masses which appear in the Lagrangian have no physical consequences, because the phases of the masses of Kaluza-Klein $ \frac {p} {R_5} + i \frac {q} {R_6}$  can eliminated by the redefinition of fields $\psi^{p, q}_{\nu 2} $. A phase in localized masses $ M_0 $  may also be eliminated by a redefinition of the fields $ \psi_{\mu 1}^0 $ and $ \psi_{\nu 1}^{p, q} $. We can then do so the following substitutions:
\begin{equation}
\frac{p}{R_5} + i \frac{q}{R_6} \rightarrow 
\sqrt{\frac{p^2}{R_5^2} +\frac{q^2}{R_6^2}} = m_{p,q}.
\label{KKMass}
\end{equation}
with a change of basis $\psi_{\mu +}^{p,q} = \frac{1}{\sqrt{2}} \left[ \psi_{\mu 1}^{p,q} + \psi_{\mu 2}^{p,q} \right]$ and $\psi_{\mu -}^{p,q} = \frac{1}{\sqrt{2}} \left[ \psi_{\mu 1}^{p,q} - \psi_{\mu 2}^{p,q} \right]$. 

In the new basis $\psi_{\mu}^{\lambda} = {\psi_{\mu 1}^0, \psi_{\mu +}^{p,q}, \psi_{\mu -}^{p,q}}$ the bi-linear terms can be expressed as:
\begin{equation}
{\cal L}_{k + m} =  \frac{1}{2} \sum_{i}  \epsilon^{\mu \nu \rho \lambda} 
\overline{\psi}^i_{\mu}\overline{\sigma}_{\nu}\partial_{\rho}\psi^i_{\lambda}
- \sum_{i,j} \psi^i_{\mu} M_{3/2~ij} \sigma^{\mu \nu} \psi^j_{\nu}
+ h.c.
\label{4dLagrangian2}
\end{equation} 
and the gravitinos mass matrix takes the form
\begin{equation}
M_{3/2} = 
\left(\begin{matrix}
m_0 & m_0 & m_0 \\
m_0 & m_0 -\delta_{kp} \delta_{lq} m_{p,q} & m_0 \\
m_0 & m_0 & m_0 +\delta_{kp} \delta_{lq} m_{p,q} \\
\end{matrix}\right) \qquad m_0 = \frac{M_0 \kappa^2}{2 \pi^2 R_5 R_6}
\label{GravitinoMassMatrix}
\end{equation}

We will now diagonalize  the mass matrix and  obtain the eigenvalues and eigenvectors. We denote by $ \Psi_m$ the eigenvector associated with the eigenvalue $ m $. It can be written in the above basis (\ref {4dLagrangian2}) as $ \Psi_m = (\Psi_m^0, \psi_{+ m}^{p , q}, \psi_{m -}^{p, q}) $. With these notations, the equations that define the vectors and eigenvalues of the mass matrix is $ M_{3/2}  \psi_{m} = m \psi_{m} $  takes the form:

\begin{eqnarray}
m_0 \left[ \psi_m^0 + \sum_{p,q \in Y} \psi^{p,q}_{m +} 
+ \sum_{p,q \in Y} \psi^{p,q}_{m -} \right]  &=& m \psi_m^0 
\nonumber    \\
m_0 \left[ \psi_m^0 + \sum_{p,q \in Y} \psi^{p,q}_{m +} 
+ \sum_{p,q \in Y} \psi^{p,q}_{m -} \right]
- m_{p,q} \psi^{k,l}_{m +}&=& m \psi^{k,l}_{m +}
\nonumber    \\
m_0 \left[ \psi_m^0 + \sum_{p,q \in Y} \psi^{p,q}_{m +} 
+ \sum_{p,q \in Y} \psi^{p,q}_{m -} \right]
+ m_{p,q} \psi^{k,l}_{m -} &=& m \psi^{k,l}_{m -} .
\label{EigenEq}
\end{eqnarray}

Some straightforward algebra leads then to the eigenvalues equation:

\begin{equation}
\sum_{p,q = - \infty}^{+ \infty} \frac{m_0}{m^2 - \frac{p^2}{R_5^2} -\frac{q^2}{R_6^2} } = \frac{1}{m}.
\label{MassEq2}
\end{equation}

We note that the double infinite sum in this equation  has a logarithmic divergence. A regularization procedure is needed and leads to a result dependent on the ultraviolet cutoff. A "truncation" of the sum  leads for the  lowest eigenvalue gravitino mass: 
\begin{equation}
\frac{1}{m_0 m} \simeq - \pi R_5 R_6.\ln \left( \Lambda^2 R^2 \right) +\frac{1}{m^2}.
\label{MassEqAprox}
\end{equation}
when taking $R \approx R_5 \approx R_6$. Retaining only the dominant terms in $M_0^2 \kappa^4 \ln \left( \Lambda R\right) /{R_5 R_6}$, we get:
\begin{equation}
\frac{1}{m} \simeq \frac{1}{m_0} + \frac{M_0 \kappa^2}{\pi} \ln\left(\Lambda R\right) .
\label{MassAprox}
\end{equation}

We see that for a small size of the extra dimensions ($R \Lambda \sim 1$), we recover the effective four-dimensional result $M_{3/2} \sim m_0$. On the other hand, for very large radius  (typically $ \ln\left( {\Lambda R}\right) > 2 \pi$),  we can  instead   have
\begin{equation}
M_{3/2} \simeq \frac{m_0}{ 2 \pi m_0^2 R^2 \ln\left( {\Lambda R}\right)} 
\label{MassAprox2}
\end{equation}
which is reduced compared to $m_0$.

We describe now the wave functions for eigenstates of the gravitinos. According to  (\ref {FourrierExpansion}) the eigenvectors of the mass matrix $ M_{3 / 2} $ (we have denoted $ \psi_{m \, \mu} $) can be written as:
\begin{equation}
\psi_{m \, \mu 1}(x^{\mu},x^5, x^6) = \frac{\kappa N e^{i \beta}}
{\sqrt{\pi^2 R_5 R_6 }} \left[ \frac{1}{\sqrt{2}} 
+ \sum_{p,q \in Y} \frac{\sqrt{2} m^2 
\cos \left( \frac{p x^5}{R_5} + \frac{q x^6}{R_6} \right) 
}{m^2 - \frac{p^2}{R_5^2} -\frac{q^2}{R_6^2}} 
\right] \chi_{m \, \mu}(x^{\mu})
\label{GravitinoEigenstate1}
\end{equation}
\begin{equation}
\psi_{m \, \mu 2}(x^{\mu},x^5, x^6) = - \frac{\kappa N m \sqrt{2} }{\sqrt{\pi^2 R_5 R_6 }} 
\sum_{p,q \in Y} 
\frac{ \sqrt{ \frac{p^2}{R_5^2} + \frac{q^2}{R_6^2} }}
{m^2 - \frac{p^2}{R_5^2} -\frac{q^2}{R_6^2}}
e^{i \alpha_{p,q}}
\sin \left( \frac{p x^5}{R_5} + \frac{q x^6}{R_6} \right) \chi_{m \, \mu}(x^{\mu}).
\label{GravitinoEigenstate2}
\end{equation}
In these expressions the field $ \chi_{m \, \mu} $ is a spinor that does not depend on extra dimensions $ (x^5, x^6) $. It is a massive spin $ 3/2$ state with mass  $ m $ as given by the equation (\ref {MassEq2}). The phases $ e^{i \beta} $ and $ e^{i \alpha_{p, q}} $ are  given by:
\begin{equation}
e^{i \beta} = \sqrt{\frac{\left| M_0 \right|}{M_0}}
,\quad 
e^{i \alpha_{p,q}} = 
\frac{\sqrt{ \frac{p^2}{R_5^2} + \frac{q^2}{R_6^2} }}{ \frac{p}{R_5} + i \frac{q}{R_6} }
e^{- i \beta}.
\label{Phases}
\end{equation}

The normalization constant $ N $ can be determined by imposing  unitary standard eigenvector $ \Psi_m = {\Psi_m^0, \psi_{m 1}^{p, q}, \psi_{m 2 }^{p, q}} $:
\begin{equation}
N = \left[ 
\sum_{p,q = - \infty}^{+ \infty} \frac{2 m^2}
{\left(m^2 - \frac{p^2}{R_5^2} -\frac{q^2}{R_6^2}\right)^2} 
- \frac{m}{m_0} 
\right]^{-1} .
\label{NormConst}
\end{equation}

One can check explicitly that  the wave functions (\ref {GravitinoEigenstate1}) and (\ref {GravitinoEigenstate2}) are solutions of the equations of motion of the gravitinos in six dimensions.

The divergence in the tree level computation (\ref{MassRunningSolut}) of the gravitino brane mass 
 arises here because of the $\delta(x_5)\delta(x_6)$ singularity due to the zero brane thickness limit. It
shows that the field theory considered here is not a valid description of the physics in the UV
as the internal structure of the brane cannot be neglected. This behavior is well known, it has been encountered in \cite{Antoniadis:1993jp} ( see also \cite{Giudice:1998ck}), and it was shown that in a  fundamental theory, as in string models, it is finite, regularized by an effective thickness \cite{Antoniadis:1993jp,Antoniadis:2000jv}.

The sensitivity to cut-off scale of the theory can be interpreted as a classical running of the mass parameter between the cut-off and the compactification scales, and it can be re-summed. It was studied in six-dimensional models with 
orbifold compactifications as a tree level renormalization of brane coupling constants \cite{Goldberger:2001tn}. While these properties have been discussed 
for particles of spin $0$ and $1/2$, here we can generalize this phenomenon for a spin $3/2$ gravitino with brane localized masses. The logarithmic divergence in (\ref{MassAprox}) can be absorbed by defining a bare coupling $M_0$ is replaced by the renormalized coupling $M_{0}^{ren}$:
\begin{equation}
\frac{1}{M_0} = \frac{1}{M_{0}^{ren}} - \frac{M_0 \kappa^4}{2 \pi^3 R_5 R_6} \ln\left(\Lambda R\right).
\label{MassRen}
\end{equation}
which implies the following running for the gravitino brane mass:
\begin{equation}
\mu \frac{d}{d \mu} M_0(\mu) = \frac{\kappa^4}{2 \pi^3 R_5 R_6}
\left[ M_0(\mu) \right] ^3.
\label{MassRunning}
\end{equation}
This   has the solution:
\begin{equation}
 M^2_0(\mu) = \frac{ M^2_0(\mu_0)}
{1 - \frac{\kappa^4}{\pi^3 R_5 R_6} M_0^2(\mu_0) 
\ln\left( \frac{\mu}{\mu_0} \right) } .
\label{MassRunningSolut}
\end{equation}

If $M_0$ is positive then equation (\ref{MassRunningSolut}) shows that 
the mass $M_0$ increases in the UV and would reach a Landau singularity at 
$ \mu = \mu_0 \exp \left[ \frac{\pi^3 R_5 R_6}{\kappa^4 M_0^2(\mu_0)} \right]$.

\section{Conclusions}

Our interest in this work, is a situation where spontaneous supersymmetry breaking is  described in a higher dimensional theory. The  presence of many localized objects (branes) coupled to bulk fields forces on the latter specific boundary conditions.  When the bulk wave functions have to interpolate between the different boundary conditions, as in the Scherk-Schwarz mechanism, this leads to supersymmetry breaking. We try to formulate these multiple boundary conditions as a system of spins, forced to point in peculiar directions by a constraining field  $ \vec{H}_i$. The amount of supersymmetry breaking is measured by the departure from the total alignment of all the  spins. The problem of building a particular configuration comes back to find the appropriate field  $ \vec{H}_i$,  and then, a microscopic realization of it.

The relevance of this picture requires that the fundamental scale, $\kappa \sim M_s^{-1}$, the branes separation distance $d_i$, and the compactification radius $R \sim M_c^{-1}$ to be separated as $\kappa \ll d_i\ll R$. A non-exhaustive set of examples is given by:
\begin{itemize}
\item the very large extra dimensions as introduced by \cite{ADD} responsible of the weakness of the strength of four dimensional gravitational interactions; these have compactification scales as low as $M_c \sim 10^{-4} -10^{7}$ eV. Along these, we can suppose the presence of 3-branes separated by typically $d_i\sim$ TeV$^{-1}$ distances. The observable world lives on some of these 3-branes or  on higher branes stretched between them.

\item The so-called  "large volume" scenario \cite{Intermediate}, with a fundamental scale in the intermediate  energies $M_s\sim 10^{11}$GeV,  with electroweak scale compactification radius $M_c \sim M_w$ , where the branes are separated by distances smaller than TeV$^{-1}$. 

\end{itemize}

Our formulation aims to study the phase space of a number  of branes located at well separated points in extra-dimensions, to discuss the "landscape" of configurations that break supersymmetry, with estimate  of its size. Given such a configuration, one needs  to explain the origin of the constraint $\vec{H}_i$,  as well as  information about the microscopic details at scales below the inter-branes distance which is relevant for many phenomenological issues.

Finally, we would like to comment on the possibility of embedding such a scenario, with discrete values for the spins  $\vec{S_i}$, in a string theory framework. In the very early history of the internal space,  one can imagine that  some   cycles shrink to a very small size. If a cycle carries some (quantized) Ramond-Ramond flux, this might give birth to a (stack of) D-brane(s) at this point (see for example \cite{Heckman:2007ub}). Different shrinking cycles can be located at points separated by potential barrier (due to wrapping factors for example) which make them potential wells where D-brane are located. The latter are driven there in order to minimize their energy contribution through the wrapping effect \cite{Giddings:2001yu}. A construction of compactifications with both branes and anti-branes exist, see for example \cite{Dabholkar:2001gz}. However, there is an issue of the stability of the configuration, either by annihilation between the brane and anti-brane, or due to decompactification pushing them infinitely away from each other. Often, studies of non-supersymmetric branes in toroidal compactifications ignore  the instability problem, having another aim as to  try to find exact conformal field theory descriptions of systems of brane-anti-branes. Most recent  studies have  concentrates on trying to find a meta-stable configuration of a single anti-brane (negatively charged) in a background with positive charge Ramond-Ramond flux (see \cite{Klebanov:2000hb}). Assuming that one can explain why the anti-brane appears at that point, and then accounts correctly for the back reaction on the backgrounds, it remains to check that the antibrane will not annihilate too quickly with part of the fluxes such as discussed in \cite{Kachru:2002gs}.    Building meta-stable string backgrounds with broken supersymmetry, even with a minimal number of branes, remains an intersecting open issue.

The main purpose here is to try instead to have a description where the whole complex system is parametrized by an effective Hamiltonian as a spin system in a bottom-up approach to the brane models constructions. The branes play here a role  similar to the  one of atoms in solid state physics, and some of the machinery of spin systems could  be applied.

\section*{Acknowledgments}

I  thank my former student C. Moura for collaboration on some parts of this work in its early stage, A. Dabholkar, B. Dou\c{c}ot and Y. Oz for useful discussions. This work is supported in parts by the European contract "UNILHC" PITN-GA-2009-237920.

\end{document}